\documentstyle[12pt,aaspp]{article}

\begin{document}

\title{Magnetohydrodynamic Turbulence in Accretion Disks}

\author{Ethan T. Vishniac}

\affil{Department of Astronomy, The University of Texas at Austin\\
Austin, Texas 78712 U.S.A.}

\begin{abstract}
Recent work on the structure of magnetic fields in a turbulent medium
gives predictions for the properties of the magnetic flux tubes as
a function of the Mach number and scale of the turbulence, and the
resistivity and viscosity of the fluid.  Here I discuss the implications
of this work for accretion disks.  I show that although accretion disk flux
tubes
are usually almost completely evacuated, they are nevertheless
less buoyant than previous estimates have suggested.  I also note that
vertical magnetic flux tends to be ejected from the outer edge of
accretion disks, so evidence
for continued magnetic activity in such systems should be interpreted
as supporting the existence of dynamo activity.
\end{abstract}


\section{INTRODUCTION}

Turbulence and magnetic fields are both topics of morbid curiousity
in astrophysics.  In that context they are usually seen as poorly
understood, undoubtedly real phenomena that can be used as part of
an explanation of last resort, i.e. when all calculable models have
been disproven.  Consequently both these phenomena, together and
separately,  have been used
in constructing models of angular momentum transport in accretion
disks, another process of indisputable reality whose nature is
obscure.  Notwithstanding this troubled history, I will present in
this paper a summary of my recent work on magnetic fields in turbulent
media and explore its implications for accretion disks.

I have three basic reasons for pressing ahead with such an unpromising
topic.  First, the structure of magnetic fields in accretion disks
determines the rate at which magnetic flux is lost from the disk.
This implies that one can get a variety of rates depending on one's
model for the magnetic field structure, but it also implies that any
physically well motivated model can have interesting, and possibly
unique,  implications.  Second, magnetic field instabilities represent
a mechanism guaranteed to move angular momentum outward and matter
inward (Balbus \& Hawley 1991), which is not true for many of the instabilities
suggested as the basis for dissipation in accretion disks.  It is
therefore critical to explore the nature of the turbulence resulting from
these instabilities.  Third, vertical magnetic fields, entrained in accretion
disks, are widely believed to be responsible for driving violent outflows,
especially jets, from a wide variety of accretion disks.  The radial
transport of vertical magnetic flux can't be understood without examining
the nature of magnetic fields in turbulent disks, and the global structure
of accretion disks fields hinges on this issue.

Before I begin I need to summarize the relevant features of accretion
disks.  The single most important point is that they are luminous
due to the conversion of orbital energy into heat.  This implies an
outward flux of angular momentum.  Such a flux would follow from the
existence of some local viscosity, but it would have to be much larger
than the viscosity implied by microscopic processes.  The usual solution
is to invoke an effective viscosity due to collective processes which
is $\nu\equiv\alpha h c_s$, where $h$ is the disk height, $c_s$ is the
sound speed, and $\alpha$ is an arbitrary constant of order unity
(Shakura and Sunyaev 1973).  For a thin disk with no self-gravity, i.e. a
`Keplerian
disk', we have a rotational frequency $\Omega(r)\propto r^{-3/2}$,
a disk height $h\sim c_s/\Omega$, and, by definition, $h\ll r$.
In this case the differential rotation in the disk plus the assumed
effective viscosity leads to an inward flux of mass given by
\begin{equation}
\dot M\sim\alpha \Sigma h^2\Omega,
\end{equation}
and a radiative flux from the disk surface of
\begin{equation}
{}F_{radiative}\sim \dot M\Omega^2.
\end{equation}
Attempts to model the outbursts for dwarf novae and X-ray transients have
led to the conclusion that $\alpha$ is probably not a constant, but
a function of local conditions (Cannizzo 1994 and references therein).
Assuming that $\alpha$
goes as $(h/r)^n$, where $n$ is a constant of order unity, gives a
reasonable fit to the observations.

In \S II of this paper I will summarize my recent work on the distribution
of magnetic fields in a turbulent medium.  In \S III I will discuss the
implications of this work for magnetic buoyancy in disks and how this
leads to a direction connection between dynamo growth rates in disks and
the appropriate value of $\alpha$.  In \S IV I will draw some general
conclusions and point the way towards future progress on this topic.

\section{MAGNETIC FIELDS IN TURBULENT FLUIDS}

The material in this section is a synopsis of Vishniac (1995a).
The basic feature of this model is that the magnetic field in
a high $\beta$ fluid, i.e. one in which the magnetic field pressure
is small compared to other sources of pressure, is spatially intermittent.
Most of the magnetic flux is contained in flux tubes, whose radii are
much smaller than the scale of curvature for the field.  This is
not a novel suggestion.  In fact, it is about what one would guess from
examining the magnetic field in the photosphere of the Sun.  It does
raise the question of how such flux tubes form.  Why should the
magnetic field and the gas spontaneously separate from one another?
The mechanism I have proposed is a process I call turbulent pumping.
If we consider an isolated flux tube in a turbulent medium, then
as long as the flux tube is flexible enough to respond to the hydrodynamic
forces exerted by the surrounding fluid then it will undergo stretching
at a rate roughly equal to the shearing rate on the scale of curvature
of the flux tube.  This lowers the linear density of matter in the
flux tube at the same rate.  By the time the flux tube length has
doubled it will be twisted by the surrounding flow in such a way that
it will intersect itself, or a whole set of neighboring flux tubes.
This will result in formation of a set of closed loops which will shrink
down to dissipative scales and vanish, thereby maintain a constant
flux tube length in the turbulent fluid.  In this way matter is removed
from the magnetic flux tubes at a rate which is dependent only on
the properties of the turbulent medium and not at all on the specific
resistivity of the fluid.   In a stationary state this loss of matter
from the flux tubes is balanced by ohmic diffusion of the charged
particles onto the field lines, a process which becomes extremely
slow as the conductivity of the fluid increases to astronomical values.
The consequence is the appearance of flux tubes whose internal gas density
can be far below that of the surrounding fluid.  In the Sun
flux tubes will be largely evacuated only near the top of the solar
convection zone, whereas in accretion disks flux tubes will be almost
empty whenever the disks are largely ionized.

This process of turbulent pumping depends on several conditions. First,
there can be little or no turbulent diffusion of matter into the
flux tubes.  Otherwise the mismatch between the mass loss driven
by collective processes (flux tube stretching and the creation of
closed loops) and mass loading driven by ohmic diffusion will disappear.
Preliminary work indicates that this condition will be satisfied
whenever the Alfv\'en speed in the flux tubes is significantly
greater than the turbulent velocity outside.  In other words, this
condition is satisfied self-consistently when turbulent pumping is
effective.  The transition from a diffuse field to one contained in
flux tubes is not yet understood.  Second, reconnection must be
efficient, in the sense of allowing the magnetic field to rearrange
its topology in less than an eddy turn over time.  Once again, this
condition is met self-consistently in the flux tube model.  This
result assumes the Sweet-Parker rate for reconnection, which is generally
considered to the slowest reasonable estimate for reconnection rates.
Third, turbulent pumping relies on the notion that closed loops whose
radii are less than a typical eddy size will tend to shrink to dissipative
scales quickly, thereby unloading their entrained mass into the surrounding
plasma.  This also follows from the physics implicit in the flux tube
description of the magnetic field, although the loops tend to shred
as they shrink, thereby creating a more diffuse, but weaker and less
organized, component to the magnetic field.

What does this pumping lead to?  If the resistivity is large enough that
particles can diffuse to the center of a flux tube in less than one
eddy turnover time, then the flux tube radius is
\begin{equation}
r_t\approx \left({\eta\over kV_l}\right)^{1/2},
\end{equation}
where $\eta$ is the resistivity, $V_l$ is the turbulent velocity on
the scale $l$ on which the flux tubes are bent, and $k$ is the corresponding
wavenumber, $k=2\pi/l$.  In this limit the magnetic pressure in a flux
tube has a gaussian profile.  When the resistivity is sufficiently small
the flux tubes will become largely empty, i.e. the magnetic field inside
the flux tube, $B_t$, is given by
\begin{equation}
{B_t^2\over 8\pi}=P,
\end{equation}
where $P$ is the pressure of the surrounding fluid.  In this limit each
flux tube will have a skin depth of width $(\eta/kV_l)^{1/2}$ surrounding
a hollow core.

None of this tells us to how decide what $r_t$ or $l$ should be in a given
situation.  For this we need to understand the forces between flux tubes, since
these forces will determine the distribution of flux elements in the turbulent
fluid.  Since in this picture the magnetic flux is confined to the interiors
of the flux tubes the forces between them will be hydrodynamic and stem from
the turbulent wakes created as the fluid moves past the semi-rigid flux tubes.
The most obvious effect (cf. Parker 1979 \S8.9) is the attraction between flux
tubes when one lies
downstream from the other.  This is simply due to the fact that the downstream
flux tube feels a reduced turbulent drag since the momentum flux around it
is reduced by $\rho V_l^2 r_t/w(r)$, where $\rho$ is the fluid density, $V_l$
is
the fluid velocity relative to the flux tubes, $r_t$ is the typical flux tube
radius, and $w(r)$ is the width of the turbulent wake at a distance $r$
downstream
from the leading flux tube.  This attraction is analogous to mock gravity, in
that
it stems from the ability of neighboring tubes to block statistically isotropic
repulsive forces in the environment.  It is less well known that one expects
neighboring flux tubes whose separation is more or less perpendicular to the
ambient flow to repel one another.  This is known experimentally (Zdravkovich
1977,
Gu, Sun, He, \& Zhang 1993)
since no adequate analytical treatment of the near-field turbulent flow is
available.  Unlike the shielding effect this repulsion depends critically on
the
nature of the flux tube wakes.  When the wakes are purely laminar and stable
there is an attractive force, which is the basis for previous claims that
flux tubes embedded in a turbulent flow always attract one another (Parker 1979
\S8.9).
The transition to an unstable wake, capable of producing repulsion, occurs when
\begin{equation}
{V_l r_t\over \nu}> \sim \pi^3.
\end{equation}
In other words, when the Reynolds number {\it on the scale of the flux tube
radius}
exceeds a critical value which lies in the range of 30 to 40.  When this
criterion
is not satisfied flux tubes will aggregate.  At higher Reynolds numbers the
magnetic
field will be broadly distributed through the fluid in the form of discrete
flux tubes
that maintain their separate identities through a rough equilibrium between
attractive
and repulsive interactions.  This qualitative difference is significant, since
this
criterion is almost always satisfied in astrophysical objects, and not (yet)
satisfied in
numerical simulations.

Neglecting viscosity, which is reasonable in stars and accretion disks, I have
used
these points to construct a simple model of the magnetic field structure in a
turbulent conducting fluid.  A detailed discussion is given in Vishniac
(1995a).
Here I simply quote the relevant results.  First, if the magnetic field energy
density is
comparable to, or less than, the turbulent energy density, there exists some
scale
$l$ such that
\begin{equation}
{B_t^2\over l}\sim \rho {V_l^2\over r_t}.
\label{eq:tense}
\end{equation}
The left hand side of this equation is the force per volume in the flux tube
exerted by magnetic tension.  The right hand side is the force per volume
exerted
by turbulent drag from the ambient fluid.  Their rough equality defines the
scale
$l$ as the scale of curvature for a typical flux tube.  In typical turbulent
cascade $V_l^2l$ is a sharply increasing function of $l$.  Consequently, on
scales
larger than $l$ the magnetic field is almost passively advected.  On smaller
scales the flux tubes are almost rigid.

Second, in order to balance the time-averaged attractive and repulsive forces
between flux tubes, it is necessary to suppose that on all scales less than $l$
the number of flux tubes within a radius $r$ of a given flux tube, $N_r$,
satisfies the condition
\begin{equation}
N_r r_t\sim r.
\end{equation}
This defines a fractal distribution of dimension one which extends from the
flux tube radius up to the scale $l$.

I can combine these results to get an expression for the average magnetic
field energy density.  This average is well-defined only on scales larger than
$l$.  On that scale I obtain
\begin{equation}
\langle B^2\rangle \sim N_l B_t^2 \left({r_t\over l}\right)^2,
\end{equation}
or
\begin{equation}
\langle B^2\rangle \sim B_t^2 {r_t\over l}.
\end{equation}
Combining this with equation (\ref{eq:tense}) we obtain
\begin{equation}
\langle B^2\rangle \sim \rho V_l^2.
\end{equation}
In other words, the scale of curvature for the flux tubes is the
scale of equipartition between the mean square magnetic field and the
average turbulent energy density.

\section{BUOYANCY AND DYNAMOS IN DISKS}

Given this specific model for the structure of a magnetic field in a turbulent
medium it is possible to discuss the systematic motion of a magnetic field
in an accretion disk.  I begin by noting that the rate at which a flux
tube will rise due to buoyant forces is given by the balance between
turbulent drag and the buoyant acceleration.  For a flux tube this gives
\begin{equation}
\Delta\rho (\pi r_t^2) g\sim \rho V_b V_l r_t,
\end{equation}
where $\Delta \rho$ is the density deficit inside the flux tube, $g$ is the
local
gravitational acceleration, and $V_b$ is the buoyant velocity.  The left
hand side of this equation is the buoyant force per unit length.  The right
hand side is the turbulent drag, assuming that $V_b\ll V_l$.  In what
follows I will assume that the magnetic field is in equipartition with
the turbulence and write $V_T$ instead of $V_l$.  Assuming that the flux
tube is small enough to be in good thermal contact with the surrounding
medium, which is usually reasonable, the fractional density deficit
$\Delta\rho/\rho$
is roughly the ratio of the magnetic pressure in the flux tube to the ambient
pressure.  This implies that
\begin{equation}
\left({B_t^2\over 8\pi P}\right)\rho (\pi r_t^2) g\sim \rho V_b V_l r_t,
\end{equation}
or
\begin{equation}
V_b\sim\left({\rho g\over P}\right) L_T V_T\sim {L_T\over l_p} V_T,
\end{equation}
where $l_p$ is the local pressure scale height and I have used the condition
that the
radius of curvature of the
magnetic field lines is $L_T$.  Note that if the magnetic field energy is
below equipartition then I need to replace $L_T V_T$ with the appropriate
$l V_l$, implying a slower buoyant rise.  For stellar convective turbulence
$l_p\sim L_T$ and magnetic flux will rise at substantial fraction of the local
turbulent velocity once the magnetic field reaches equipartition with the
turbulence.

This result cannot be extended to accretion disks.  There the most plausible
source of turbulence is magnetic field instability first described by
Velikhov (1959) (see also Chandrasekhar 1961) and applied to accretion disks by
Balbus and Hawley (1991).   Here I invoke the description of the saturated
state of this instability for a large scale azimuthal field embedded in an
accretion disk given in Vishniac and Diamond (1992).  The dominant eddies
will be those characterizing the fastest growing mode, for reasons explained
in that paper.  The instability will saturate in a turbulent state
characterized
by a typical turbulent velocity comparable to the Alfv\'en speed, i.e.
$V_T\sim V_A$.  The eddy size will be $L_T\sim V_A/\Omega$, where $\Omega$
is the local rotational frequency.  This gives rise to an effective
viscosity and diffusion coefficient of order $L_TV_T\sim V_A^2/\Omega$
or $(V_A/c_s)^2 h^2\Omega$, where $h$ is the disk thickness and I have
used the relationship $c_s\sim h\Omega$, which applies to thin,
non-self-gravitating
accretion disks.  The dimensionless viscosity of the disk, due to magnetic
field stresses, is just $(V_A/c_s)^2$.

What does this imply about flux tubes in accretion disks?  It can be shown
(Vishniac 1995b) that flux tubes in hot disks are in the ideal fluid regime,
i.e. they are almost completely empty.  In this case
\begin{equation}
r_t\sim L_T\left({V_T\over c_s}\right)^2\sim \left({V_A\over c_s}\right)^3
h\sim
\alpha^{3/2} h.
\end{equation}
The buoyant velocity is
\begin{equation}
V_b\sim {L_T\over l_p} V_T\sim {V_A/\Omega\over h} V_A\sim {V_A^2\over
c_s}\sim\alpha c_s.
\end{equation}
Since the loss rate for magnetic flux is just $V_b/h\sim\alpha \Omega$ this
implies that the
azimuthal magnetic field escapes from the disk at a rate which is comparable to
the
thermal relaxation rate for an optically thick disk.  I note in passing that
this
flux loss rate is smaller, by a factor of $V_A/c_s$, than estimates based on
the Parker
instability, which is normally taken to imply a buoyant velocity $\sim V_A$.
The reason for this discrepancy is that the Parker instability is strongly
suppressed by the Balbus-Hawley instability (Vishniac and Diamond 1992).
In a stationary state the magnetic field must be regenerated by some dynamo
process
so that the dynamo growth rate balances the buoyant flux losses, i.e.
\begin{equation}
\Gamma_{dynamo}\sim {V_b\over h}\sim \alpha\Omega,
\end{equation}
or
\begin{equation}
\alpha\sim {\Gamma_{dynamo}\over \Omega}.
\end{equation}

When radiation pressure is large and electron scattering dominates the opacity,
a situation normally encountered in the inner regions of AGN disks, the flux
tube
properties change significantly.  In this case the magnetic pressure in the
flux tubes is limited by the ambient gas pressure, since the ambient photons
can
diffuse into the flux tubes on a very short time scale.  This does not affect
the efficiency of turbulent pumping, with the consequence that the flux tubes
are larger than one would expect in the ideal fluid regime, but are still
evacuated.
This gives a modified expression for the flux tube radius, i.e.
\begin{equation}
r_t\sim L_T\left({V_T\over c_{s,gas}}\right)^2\sim h\left({V_A\over
c_s}\right)^3
{P\over P_{gas}}\sim\alpha^{3/2} h{P\over P_{gas}}.
\end{equation}
This leads to an enhanced buoyancy so that
\begin{equation}
V_b\sim g\left({\rho V_T^2 L_T\over P_{gas}}\right) {1\over V_T}\sim{P\over
P_{gas}}
\alpha c_s.
\end{equation}
Consequently, for a given dynamo growth rate, balancing magnetic flux
generation
with buoyant losses I get
\begin{equation}
\alpha\sim\left({\Gamma_{dynamo}\over\Omega}\right)\left({P_{gas}\over
P}\right).
\end{equation}
In other words, as long as the dynamo growth rate is independent of the
magnetic energy
density the implied disk viscosity will scale with the gas pressure, rather
than the
total pressure.  Of course, this may imply that other angular momentum
transport mechanisms,
normally dominated by magnetic stresses, become important.

What are some possible disk dynamos?

The equilibrium state of the azimuthal magnetic field is determined by the
balance between
dynamo activity and vertical buoyancy.  However, one can also imagine that a
typical
accretion disk will have a large scale vertical magnetic field, if for no other
reason
than the fact that such a field is likely to be accreted along with matter
added to
the outer edge of the disk.  Clearly vertical buoyancy is irrelevant to the
evolution
of this field.  Moreover, since the field lines cross the disk in concentrated
flux
tubes, and spread out above and below the disk, the tension due to strong
bending
of the external field lines is negligible.  Nevertheless, there are
two effects which will tend to move vertical field lines
outward.  First, if the field lines are bent radial by some total angle
$2\theta$
as they cross the disk, then turbulent diffusion through the disk will tend to
combine radial field lines of opposite polarity, moving the point at which the
field lines cross the disk outward at a rate of roughly $\alpha c_s\tan\theta$
(Van Ballegooijen 1989).  If the magnetic field curvature is determined only by
large
scale stresses than $\theta\sim h/r$ and this velocity is comparable to the
inward flow of matter in the disk.  Consequently it will be difficult to
determine the direction of drift for the magnetic field.  Of course, if the
concentration of magnetic field in the inner disk increases, then the global
stresses will increase and the disk will stop accreting vertical flux
regardless.
In addition, if the field is responsible for driving a wind or jet (cf. Shu et
al.
1994 and references therein) then it will tend to bend sharply near the
disk which will move the field outward.

Second, the flux tubes containing the vertical flux will be subjected to
radial buoyancy forces (Parker \& Vishniac 1995).  Each flux tube will have an
associated
energy due to its displacement of matter and associated pressure.  This energy
has
two (usually) comparable parts.  The first is due to the surrounding pressure
and is
roughly equal to $\Delta P L \pi r_t^2$, where $L$ is the length of the flux
tube, and
$\Delta P$ pressure contributed by the magnetic field in the flux tube.  The
other
is due to the displacement of matter which could otherwise settle to the
disk midplane.  This term is of order $\Delta \rho (h\Omega)^2 L\pi r_t^2$,
where
$\Delta\rho$ is the density deficit in the flux tube and is of order $\rho$
under
normal circumstances in a hot disk.  For a gas pressure dominated disk
the two are comparable and roughly equal to $B_t^2r_t^2 L$.  For a radiation
pressure
dominated disk the gravitational term dominates is larger by a factor of
$P/P_{gas}$.
One would get the same effect by replacing the $\Delta P$ in the pressure
contribution
with $P$ rather than $P_{gas}$.
If the energy associated with a flux tube is $U_t$ then the consequent radial
drift
velocity is obtained by equating the turbulent drag with the radial gradient of
$U_t$, or
\begin{equation} \rho V_TV_b r_t \sim -{\partial_r U_t\over L}\sim -P
r_t^2\partial_r (\ln U_t).
\end{equation}
Consequently,
\begin{equation}
V_b \sim -\alpha c_s h {P\over P_{gas}}\partial_r (\ln U_t),
\end{equation}
which for $\partial_r(\ln U_t)\sim r^{-1}$ implies a radial drift velocity
which
is larger than the inward accretion velocity by a factor of $P/P_{gas}$.  The
direction of the drift depends on the sign of $\partial_r(\ln U_t)$.  Since
a single flux tube can break apart, or combine, in the course of its radial
drift,
we need to evaluate this derivative under the constraint that the magnetic flux
remains fixed, or that the area goes as $P_{gas}^{1/2}$.  The length of the
flux tube will exceed $h$, since each tube actually crosses the disk in a
random
walk.  We will assume that $L\propto h\alpha^{-1/2}$.  Subject to these
constraints
we find that
\begin{equation}
V_b \sim -\alpha c_s h {P\over P_{gas}}{1\over2}\partial_r \left({P\dot M
c_s\over P_{gas}\alpha}
\right).
\end{equation}
If $\dot M$ is constant then this will almost give a strong outward buoyancy to
the flux
tubes, which will clearly dominate over accretion when $P\gg P_{gas}$.  In the
event
that the inner disk is unstable and $\dot M$ varies with $r$ there will still
be a
averaged outward flow.  The evidence for magnetic activity in accretion disks,
especially
in AGN, must be read as evidence for large scale dynamo activity in these
disks.

What are the prospects for a reasonable theory of dynamo activity in disks?
There
is an extensive literature on this topic, too extensive to summarize here.  My
own
view is that there are only a few processes which we can be reasonably sure
exist and
which may dominate in real accretion disks.  All of them rely on the notion
that
shearing of a radial field provides for efficient generation of a large scale
azimuthal
field.  The tricky step is to understand how the azimuthal field component
regenerates
the radial field.  One of the most popular notions is that magnetic buoyancy
produces
turbulent motions with a preferred helicity, which close the cycle by twisting
the azimuthal field lines into radial field lines (Galeev, Rosner, \&
Vaiana 1979).  However, this relies on
using the Parker instability, which has a large azimuthal wavenumber.  Since
modes
with a large $k_\theta$ and a slow growth rate will be suppressed by the
Balbus-Hawley
instability, this mode is not expected to exist in real accretion disks
(Vishniac
and Diamond 1992), nor is it seen in numerical simulations (Brandenburg et al.
1995).
Another idea is that internal waves, excited near the outer edge of the disk
via
tidal forcing (Goodman 1993), will fill the disk and drive a dynamo with a
growth
rate $\sim (H/r)^{3/2}$ (Vishniac, Jin, and Diamond 1990, Vishniac and Diamond
1992).
This implies a dimensionless viscosity of $\sim (H/r)^{3/2}(P_{gas}/P)$.
{}Finally,
Balbus and Hawley (1991) have claimed that the Balbus-Hawley instability will
lead to a local turbulent dynamo which will saturate near equipartition between
the field and ambient pressure, leading to an $\alpha$ of order unity.
Something
like this is seen in numerical simulations (e.g. Brandenburg et al. 1995),
although
it would appear to be inconsistent with phenomenological work on accretion
disks
(cf. Cannizzo 1994 and references therein).  On
the other hand, the numerical results are also consistent with the idea that
the
Balbus-Hawley instability is driving an incoherent dynamo according to mean
field
theory, leading to a saturation which depends on the geometry of the
computational
box (Vishniac and Brandenburg 1995).  This theory predicts an effective
$\alpha$
of order $(H/r)^2(P_{gas}/P)^6$.  The steeper scaling with $H/r$ and strong
suppression
in radiation pressure dominated environments implies that while this process
may play a role in real disks, it will only dominate in disks not subject to
significant elliptical distortions at large radii and free of significant
radiation pressure.

\section{CONCLUSIONS AND FUTURE PROSPECTS}

I can summarize my results as follows.  First, I have proposed a simple model
of flux tube
formation that is consistent with solar observations.  Second, in this picture
there
is a failure of `flux-freezing' to adequately describe the macroscopic motions
of the
field and fluid.  In general there will be some significant relative motion
between
the flux tubes and the fluid.  This suggests the possibility of mean-field
dynamos.
Third, direct numerical simulation of astrophysically realistic driven MHD
turbulence
is not currently possible.  Turbulence driven by magnetic field instabilities
can
be simulated to obtain qualitative results, but such simulations will be
quantitatively unreliable.  Fourth, accretion disks might be able to move
magnetic field lines inward, but only if their radial bending angle across the
disk
is of order $h/r$ or less.   Even in this case radial buoyancy may move them
outward.  Fifth, magnetic viscosity in AGN disks couples primarily to gas
pressure,
not radiation pressure.  This may imply that the viscosity due to hydrodynamic
effects, e.g. internal wave breaking, dominates.  Finally, given the relatively
high rate of vertical and radial buoyant magnetic flux losses, evidence for
continued
magnetic activity is disks (and stars) can only be explained by the presence of
some dynamo mechanism.

It is somewhat disappointing that the MHD turbulence model used here implies
that
numerical simulations of astrophysical turbulence are not physically realistic.
{}Fortunately, this same theory does predict scaling behavior and approximate
saturation
values for magnetic fields in the viscous regime, which is currently accessible
to
numerical simulations.  Somewhat fragmentary results from current work seem to
show the predicted behavior of the magnetic field as a function of Reynolds
number
(Vishniac 1995a).  Future tests of the theory in this regime should give us
some
confidence in its application to astrophysics, or allow us to discard it in
favor
of some alternative description.  However, until we are in position to do more
realistic MHD simulations it will not be possible to replace the approximate
results sketched here with quantitative estimates.


\begin{references}

\reference
Balbus, S.A., \& Hawley, J.F.
1991, ApJ, 376, 214

\reference
Brandenburg, A., Nordlund, A, Stein, R.F., \& Torkelsson, U.
1995, ApJ, in press

\reference
Cannizzo, J.K.
1994, ApJ, 435, 389

\reference
Chandrasekhar, S.
1961, Hydrodynamic and Hydromagnetic Stability
(Oxford: Oxford University Press)

\reference
Galeev, A.A., Rosner, R., \& Vaiana, G.S.
1979, ApJ, 229, 318

\reference
Goodman, J.
1993, ApJ, 406, 596

\reference
Gu, Z.F., Sun, T.F., He, D.X., \& Zhang, L.L.
1993, Journal of Wind Eng. and Ind. Aero., 49, 379

\reference
Park, S.J., \& Vishniac, E.T.
1995, ApJ, submitted

\reference
Parker, E.N.
1979, Cosmical Magnetic Fields, (Oxford: Oxford University Press)

\reference
Shakura, N.I., \& Sunyaev, R.A.
1973, AA, 24, 337

\reference
Shu, F., Najita, J., Ostriker, E., Wilkin, F, Ruden, S., \& Lizano, S.
1994, ApJ, 429, 781

\reference
Van Ballegooijen, A.A.
1989, in Accretion Disks and Magnetic Fields in Astrophysics,
(Dordrecht, Kluwer), 99

\reference
Velikhov, E.P.
1959, Soviet Phys. JETP Lett., 35, 1398

\reference
Vishniac, E.T.
1995a, ApJ, in press

\reference
Vishniac, E.T.
1995b, ApJ, in press

\reference
Vishniac, E.T., \& Brandenburg, A.
1995, in preparation

\reference
Vishniac, E.T., \& Diamond, P.H.
1992, ApJ, 398, 561

\reference
Vishniac, E.T., Jin, L., \& Diamond, P.H.
1990, ApJ, 365, 552

\reference
Zdravkovich, M.M.
1977, Trans. of the ASME: Journal of Fluids Eng., 99, 618

\end{references}
\end{document}